\documentclass{aa}

\usepackage{graphicx}
\usepackage{txfonts}
\usepackage{bm}
\usepackage[colorlinks,allcolors=blue]{hyperref}

\makeatletter
\renewcommand*\aa@pageof{, page \thepage{} of \pageref*{LastPage}} 
\makeatother
\usepackage{amstext}

\begin{document}

\title{USuRPER: Unit-sphere representation periodogram for full spectra}

\author{A. Binnenfeld\inst{\ref{inst1}} \and S. Shahaf\inst{\ref{inst2}} \and S. Zucker\inst{\ref{inst1}}}

\institute{Porter School of the Environment and Earth Sciences, Raymond and Beverly Sackler Faculty of Exact Sciences, 
Tel Aviv University, Tel Aviv, 6997801, Israel \\
\email{avrahambinn@gmail.com,shayz@tauex.tau.ac.il}\label{inst1}
\and
School of Physics and Astronomy, Raymond and Beverly Sackler Faculty of Exact Sciences, 
Tel Aviv University, Tel Aviv, 6997801, Israel \\
\email{sahar.shahaf@gmail.com}
\label{inst2}
}

\date{Accepted XXX. Received YYY}

\abstract{
We introduce an extension of the periodogram concept to time-resolved spectroscopy. USuRPER, the unit-sphere representation periodogram,  is a novel technique that opens new horizons in the analysis of astronomical spectra. It can be used to detect a wide range of periodic variability of the spectrum shape. Essentially, the technique is based on representing spectra as unit vectors in a multidimensional hyperspace, hence its name. It is an extension of the phase-distance correlation periodogram we had introduced in previous papers, to very high-dimensional data such as spectra. USuRPER takes the overall shape of the spectrum into account, which means that it does not need to be reduce into a single quantity such as radial velocity or temperature. Through simulations, we demonstrate its performance in various types of spectroscopic variability: single-lined and double-lined spectroscopic binary stars, and pulsating stars. We also show its performance on actual data of a rapidly oscillating Ap star. USuRPER is a new tool to explore large time-resolved spectroscopic databases such as\ APOGEE, LAMOST, and the RVS spectra of \textit{Gaia}. We have made a public GitHub repository with a Python implementation of USuRPER available
to the community, to experiment with it and apply it to a wide range of spectroscopic time series.
}


\titlerunning{USuRPER: periodogram for full spectra}
\authorrunning{A. Binnenfeld et al.}

\maketitle


\section{Introduction}
\label{sec:intro}
In two previous papers \citep{Zuc2018,Zuc2019}, we have introduced the phase-distance correlation (PDC) periodogram as a new method to detect non-sinusoidal periodicities in unevenly sampled time-series data. Essentially, for each trial period, PDC quantifies the statistical dependence between the measured quantity and the phase (according to the trial period), using the recently introduced \emph{\textup{distance correlation}}. \citet*{Szeetal2007} introduced distance correlation as a measure of statistical dependence between two quantities. The calculations involved in estimating the sample distance correlation somewhat resemble those involved in estimating the Pearson correlation, hence its name. However, it is important to note that unlike the Pearson correlation, the distance correlation is not a measure of linear dependence, but rather of general dependence.

In order to quantify the dependence on the phase, which is a circular variable (i.e.\ cyclic), we have modified the original expression of \citet{Szeetal2007}, following their original derivation, but for circular variables. As we have shown \citep{Zuc2018}, the newly introduced periodogram outperforms other methods in cases of sawtooth-like variability shapes, including also radial velocity (RV) curves of eccentric single-lined spectroscopic binary (SB1) stars. 

We have later extended the PDC periodogram to two-dimensional data \citep{Zuc2019}, and specifically, to two-dimensional astrometric data, so as to improve the detection of eccentric astrometric orbits. This generalisation demonstrated an important advantage of distance correlation over the classic Pearson correlation. The Pearson correlation involves products of the sample values of the two examined variables, which means that both of them have to be real numbers. Instead, the distance correlation involves element-wise products of two matrices that are based on the distance matrices of the two variables. As long as distances can be calculated in each of the two examined spaces, no requirement regarding the dimensionality of the two variables is therefore made. They can even be of different dimensions, as long as distance matrices can be computed.

\citet{Lyo2013} further extended the applicability of distance correlation by showing that it can be applied to variables in general metric spaces, as long as the two metrics involved are both of \lq strong negative type\rq. It is beyond the scope of this paper to delve into the definition and subtleties of strong negative-type metric spaces \citep*[first introduced in][]{Zinetal1992}, but it is still important to note that Euclidean spaces are of strong negative type \citep{Lyo2013}.

In this paper we introduce an extension of the PDC periodogram to a new domain: we propose to use it to detect general periodic variability of astronomical spectra. Perhaps the most obvious periodic variability of a stellar spectrum is that of SB1s, in which the spectra exhibit periodic Doppler shifts. The usual way to study SB1s is to cross-correlate the spectra against a template spectrum (either synthetic or observed), derive an estimate of the Doppler shifts from the location of the cross-correlation peaks \citep[e.g.][]{TonDav1979}, and then analyse the Doppler shifts in search for periodicity using conventional techniques, such as the generalized Lomb-Scargle (GLS) periodogram \citep{Fer1981,ZecKur2009}. 

Double-lined spectroscopic binaries (SB2s) exhibit a more complicated periodicity pattern because each observed spectrum is essentially a superposition of two spectra, each shifted by a different Doppler shift, and both undergo opposite RV changes. Occasionally, the cross-correlation of the spectrum against a template shows two peaks, but sometimes the two peaks blend, requiring the use of techniques to disentangle the two Doppler shifts, such as \ TODCOR \citep{ZucMaz1994}.
\citet{SimStu1994} proposed a disentangling technique that did not require early knowledge of the component spectra. However, their approach is still tailored only to SB2s.

Periodic variability of the spectrum need not necessarily be related to Doppler shifts in binary stars. Various types of stellar pulsations bring about many types of periodic variations of the spectrum, ranging from periodic temperature changes such as in Cepheids \citep*[e.g.][]{Andetal2005} to line-profile variations in non-radially pulsating stars \citep*[e.g.][]{Aeretal1992}.

In the next section we introduce the details of the calculations involved in producing the USuRPER periodogram. To demonstrate the capabilities of USuRPER, we show in Sect.~\ref{sec:ex} some test cases, both simulated and actually observed. We finally conclude in Sect.~\ref{sec:conc} with a short summary and some insights regarding applicability.

\section{Unit-sphere representation periodogram}
\label{sec:usurper}

\subsection{Fundamentals}
\label{subsec:fund}

We assume that we have time-resolved spectroscopy data of an astronomical object, comprising $N$ spectra obtained at times $\{t_i\}_{i=1}^N$. We further assume that each spectrum is essentially an array of $L$ intensities, each corresponding to a specific wavelength. For simplicity, we assume at this stage that all spectra are calibrated to the same wavelength grid, and are all measured at the same rest frame. These assumptions can later be easily relaxed by calibration and interpolation procedures that are routinely performed in astronomical spectroscopy and RV studies.

Because we are interested only in the variability of the spectrum shape (rather than the total flux), we subtract the mean value of each spectrum and normalise it by dividing with its standard deviation. As a result, the spectra, $\{\bm{\hat{f_i}}\}_{i=1}^N$, can be now considered unit vectors in an $L$-dimensional Euclidean space, that is,\ points on the unit $(L-1)$-sphere. If a periodic variability of the spectrum shape were to take place, it would therefore be manifested in a periodic motion on this unit sphere. 

We introduce here a novel kind of periodogram to look for this unit-sphere periodicity. Following our previous papers, we can construct such a periodogram by quantifying for each trial period the distance correlation between the location on the unit sphere and the phase (according to the trial period). To do this, we need to have a distance function (metric) on the unit sphere that will be of strong negative type. Such a metric can be defined by the length of the chord connecting two points on the sphere: \emph{\textup{the chord-length metric}}. This metric is of strong negative type because it is induced by the Euclidean metric of the $L$-dimensional space in which the unit sphere is embedded \citep{Lyo2013}. As we now show, this metric is very easy to compute. 

Let $\bm{\hat{f_i}}$ and $\bm{\hat{f_j}}$ be two members of the sequence of unit vectors introduced above. We denote by $\theta_{ij}$ the angle between these two unit vectors. By simple geometry, we can immediately see that the chord length between the two corresponding unit-sphere locations is given by 
\begin{equation}
\label{eq:dist}
d(\bm{\hat{f_i}},\bm{\hat{f_j}}) = 2\sin(\theta_{ij}/2) = 2 \sqrt{\frac{1-\cos\theta_{ij}}{2}} \ .
\end{equation}
Because $\bm{\hat{f_i}}$ and $\bm{\hat{f_j}}$ are unit vectors, $\cos\theta_{ij}$ is in fact the scalar product between them. In other words, it is actually the \emph{\textup{normalised correlation between the two original spectra}}, henceforth $C_{ij}$. 

Now that we have defined a distance function, it might appear that we can calculate the two required distance matrices, following \citet{Zuc2018,Zuc2019}. However, the space on which our distance function (Eq.~(\ref{eq:dist})) is defined is extremely high dimensional, and as \citet{SzeRiz2013} showed, a naive computation of the distance correlation in this case would introduce a very strong bias. They proposed instead to use an \emph{\textup{unbiased}} estimate of the distance correlation, which we introduce in the next paragraphs.

\subsection{Computation}

Similarly to \citet{Zuc2018,Zuc2019}, we define a distance matrix based on the metric we have introduced in Eq.~(\ref{eq:dist}). For each pair of spectra ($i$ and $j$), the entry in the distance matrix is
\begin{equation}
a_{ij} = \sqrt{1-C_{ij}} \ .
\end{equation}
We can safely remove the multiplicative factors appearing in Eq.~(\ref{eq:dist}) because they would later cancel out in the normalisation.

For each trial period $P$ we define a phase-distance matrix, similarly to that in previous papers:
\begin{equation}
\begin{split}
\phi_{ij} &= (t_i - t_j)\mod P \ , \\
b_{ij} &= \phi_{ij}(P-\phi_{ij}) \ .
\end{split}
\end{equation}

Now, instead of the zero-centring used in the previous papers, which leads to a biased estimator of the distance correlation,  we apply $\mathcal{U}$-centring, introduced in \citet{SzeRiz2014} in order to mitigate the bias:
\begin{equation}
A_{ij} = \begin{cases}
\begin{split}
a_{ij} - \frac{1}{N-2}\sum\limits_{k=1}^{N}a_{ik} 
- \frac{1}{N-2}\sum\limits_{k=1}^{N}a_{kj} \\
+ \frac{1}{(N-1)(N-2)}\sum\limits_{k,l=1}^{N}a_{kl} 
\end{split} 
& \text{if $i \neq j$ ,} \\ \\
0 & \text{if $i = j$ .}
\end{cases}
\end{equation}

A similar procedure is applied to obtain the matrix $B_{ij}$ from $b_{ij}$. When the $\mathcal{U}$-centred matrices are used, the unbiased estimator of the distance correlation can be computed by the expression
\begin{equation}
\label{eq:cor}
D = \frac{\sum\limits_{ij}A_{ij}B_{ij}}{\sqrt{(\sum\limits_{ij}A^2_{ij})(\sum\limits_{ij}B^2_{ij})}} \ .
\end{equation}

If prominent peaks appear in the resulting periodogram, their significance can be assessed by a permutation test. Every spectrum would then be allocated a random phase, drawn uniformly, and $D$ would be recalculated for this random allocation of phases. There would be no need to recalculate the distance matrix among the spectra, as the original phase dependence would already have been ruined by randomising the phases. By repeating the randomisation for a prescribed number of times, the sample of $D$ values can be used to obtain a threshold value corresponding to a desired level of the false-alarm probability (FAP).

\subsection{Run-time complexity}

The matrix $A_{ij}$ should be calculated only once. If the spectra are all calibrated to the same wavelength grid, each $C_{ij}$ is a simple correlation coefficient, requiring $\mathcal{O}(L)$ operations. However, because cross-correlation functions (CCFs) are routinely computed, especially in the context of RV studies, $C_{ij}$ can also be extracted from the CCF, taking into account conversion to the rest-frame velocity. CCFs usually require $\mathcal{O}(L\log L)$ operations (using fast convolution algorithms), which we henceforth use as a worst-case estimate. Therefore, calculating the matrix $A_{ij}$ and converting it into the $\mathcal{U}$-centred matrix $a_{ij}$ involves $\mathcal{O}(N^2L\log L)$ operations. The matrix $b_{ij}$ has to be calculated separately for each frequency, and then used to calculate the distance correlation (Eq.~(\ref{eq:cor})), amounting to a total of $\mathcal{O}(N^2K)$, where $K$ is the number of trial frequencies (periods). The total time complexity is therefore $\max\left[\mathcal{O}{(N^2L\log L}),\mathcal{O}({N^2K})\right],$ and it is a matter of specific implementation which of the two terms dominates. Whichever dominates, it is still a matter of quadratic dependence on the number of spectra. In future applications, this quadratic dependence may be reduced to $\mathcal{O}(N\log N)$ by using fast techniques to compute distance correlation that are now emerging \citep[e.g.][]{HuoSze2016,ChaHu2019}.

\section{Examples}
\label{sec:ex}

\subsection{Simulated SB1}
\label{subsec:SB1}

In order to simulate spectra of an SB1, we used a synthetic solar-like spectrum ($T_\mathrm{eff}=5800\,\mathrm{K}$, $[\element{Fe}/\element{H}]=0.0$, $\log g=4.5$) from the spectral library PHOENIX \citep{Husetal2013}, at a spectral resolution of $R=10000$. We have simulated a simple sinusoidal RV curve (i.e.\ corresponding to a circular orbit), with a semi-amplitude of $K=10\, \mathrm{km\,s}^{-1}$ and a period of $\text{seven}$\,days. We randomly drew $50$ epochs from a uniform distribution on an interval of $100$~days, and after shifting the spectrum according to the required RVs, we added to the spectra white Gaussian noise, at a signal-to-noise ratio (S/N) of $100$\footnote{The SNR definition we used was the ratio between the continuum flux level and the noise standard deviation. We estimated the continuum flux level by the $98$-th percentile of the flux values in the spectrum.}. The wavelength range we have used for our simulations was $4900$\,--\,$5100$\,\AA.

\begin{figure}
\includegraphics[width=\columnwidth,clip=true]{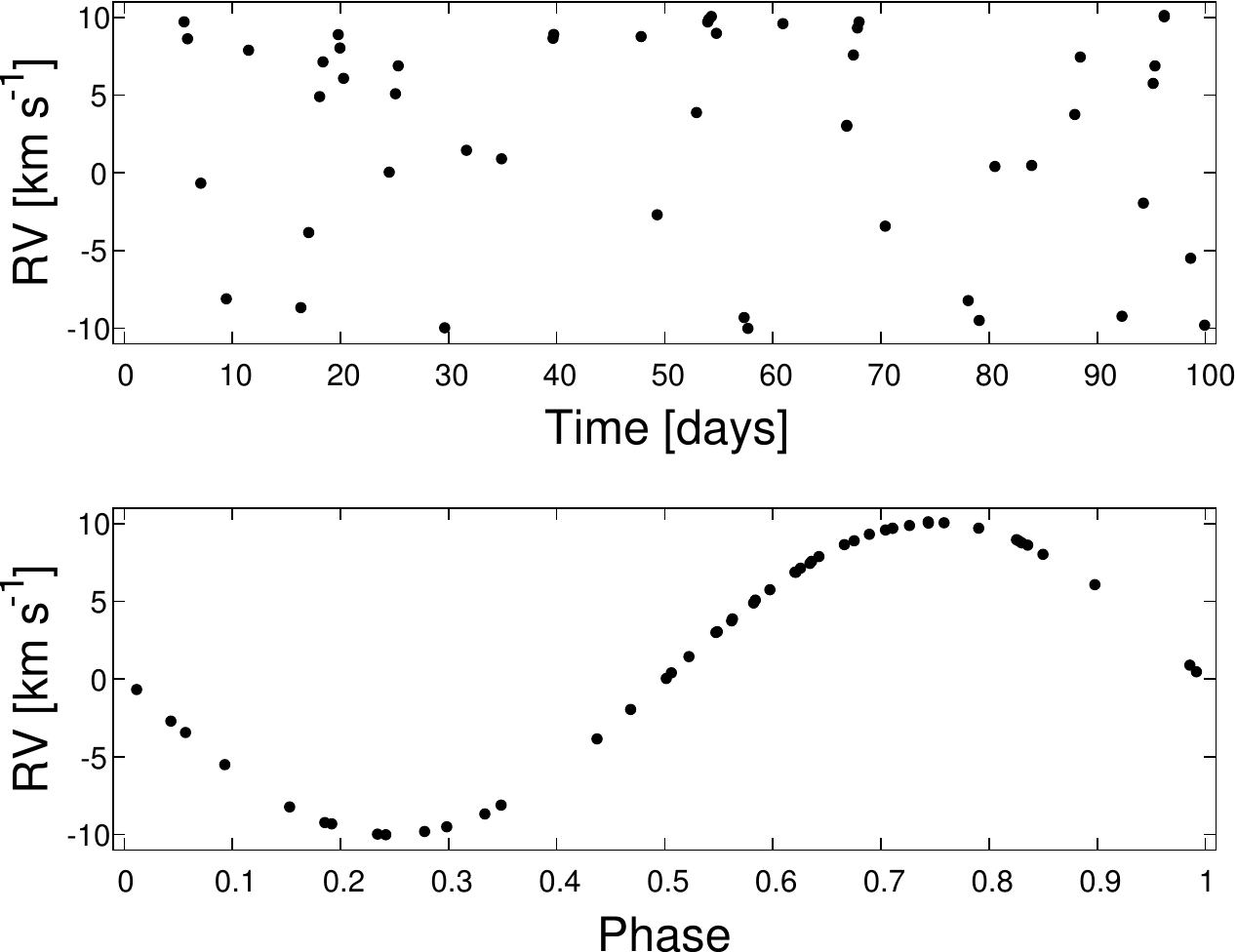}
\caption{Upper panel: Estimated RV time series based on the simulated spectra of an SB1. Lower panel: RV time series phase-folded by the known $\text{seven}$-day period.}
\label{fig:SB1RV}
\end{figure}

\begin{figure}
\includegraphics[width=\columnwidth,clip=true]{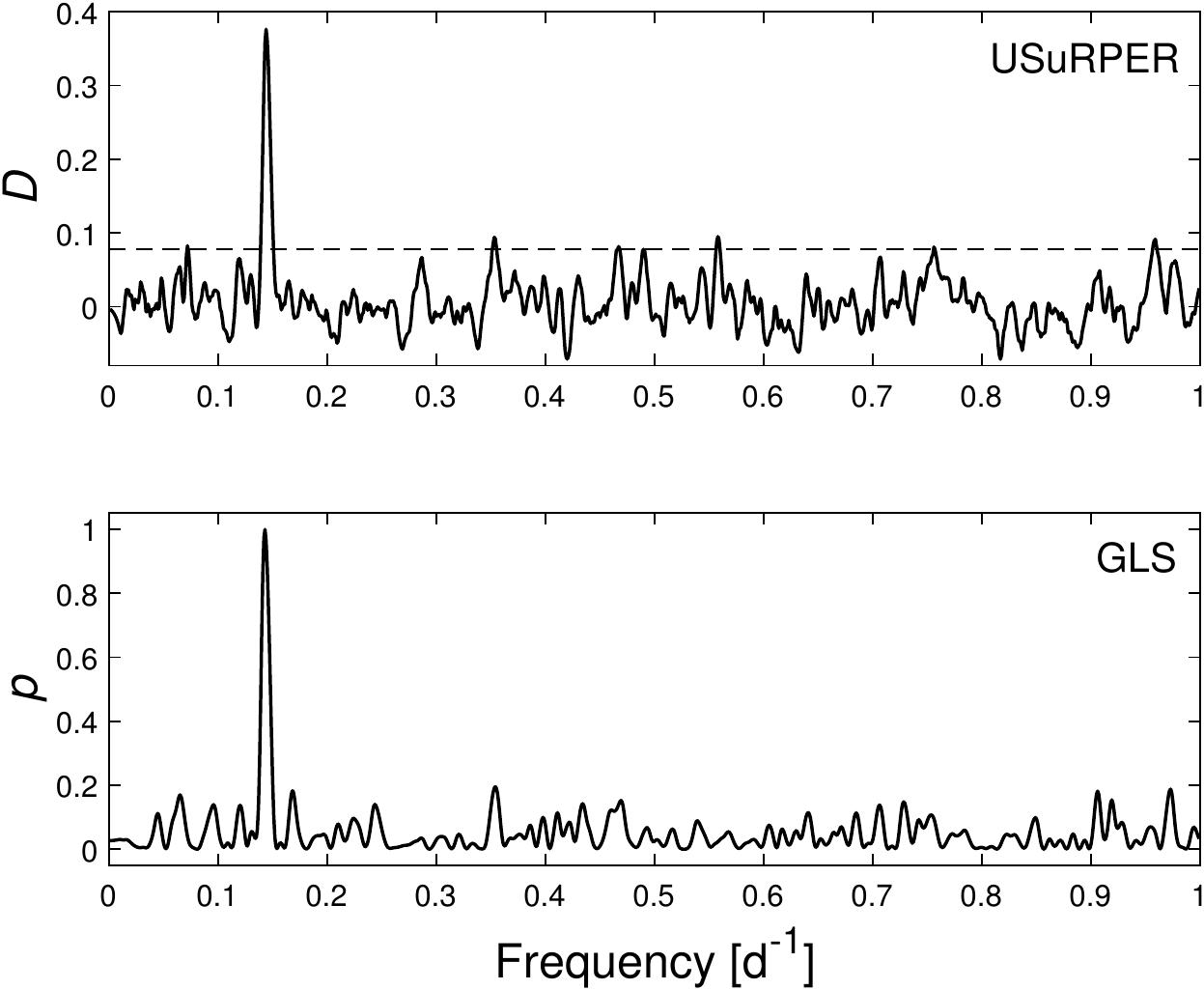}
\caption{GLS (lower panel) and USuRPER (upper panel) periodograms of the simulated SB1 whose RV are presented in Fig.~\ref{fig:SB1}. 
The GLS power and the distance correlation values of USuRPER are both normalised and therefore unitless. The dashed line in the upper panel corresponds to an FAP level of $10^{-3}$, obtained by the permutation test procedure.}
\label{fig:SB1}
\end{figure}

The common approach to analysing such data is to cross-correlate each observed spectrum against an assumed template and estimate the location of the cross-correlation peak. Fig.~\ref{fig:SB1RV} shows the resulting RV estimates thus obtained (using the PHOENIX spectrum as template). As is clearly evident from the figure, the high S/N we used in the simulation led to what seems to be a very smooth sinusoidal RV curve, with negligible scatter around the sinusoid. This very well-defined periodicity, combined with the relatively large number of samples, is also manifested in a very sharp and prominent peak in the GLS periodogram at a frequency of $1/7\,\mathrm{d}^{-1}$ (Fig.~\ref{fig:SB1}, lower panel). Because the GLS is tailored to sinusoidal periodicities, we do not expect any other kind of periodogram to outperform the GLS in this case. Moreover, when we search for periodicity in the RV data, it means that we have already assumed that the spectroscopic variability is a Doppler-shift periodicity, and not, for example, line-profile variation.

Nevertheless, it is illuminating to compare GLS to our newly introduced periodogram. The upper panel of Fig.~\ref{fig:SB1} shows the resulting USuRPER periodogram. We recall that we did not extract RVs in order to obtain this periodogram, therefore it is very encouraging that USuRPER produced such a sharp peak at the correct period. The dashed line in the plot shows the threshold value corresponding to an FAP of $10^{-3}$, leaving little doubt concerning the significance of the detected periodicity.

This example is a very simple case, with many samples and a high S/N. It still serves as a kind of sanity check, and proves that this novel approach can indeed identify at least simple periodicities.

\subsection{Simulated SB2}
\label{subsec:SB2}

\begin{figure}
\includegraphics[width=\columnwidth,clip=true]{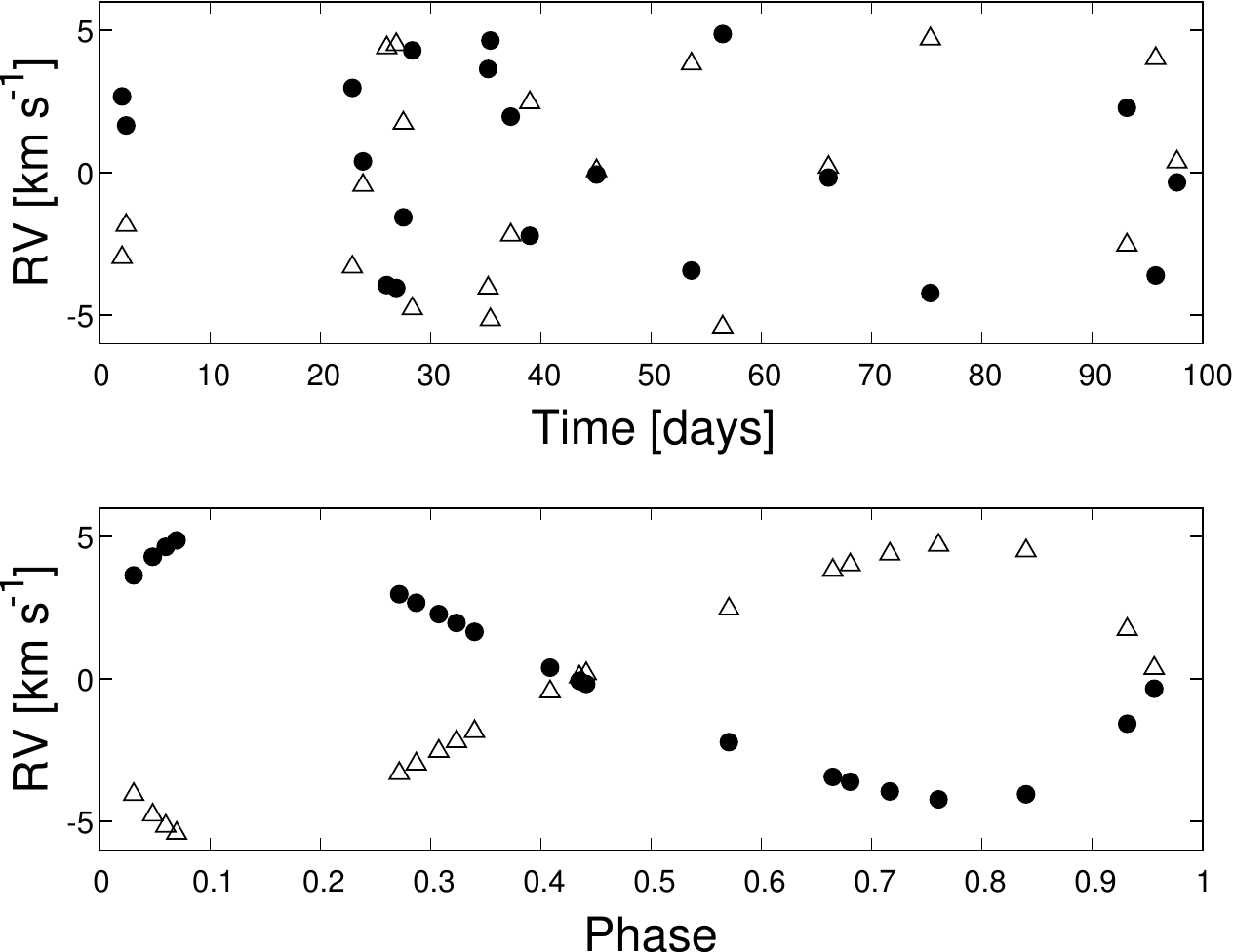}
\caption{Upper panel: RV time series used in the SB2 simulation. Filled circles mark the primary RV, and empty triangles the RV of the secondary. Lower panel: Same RV time-series phase-folded by the known $\text{seven}$-day period.}
\label{fig:SB2RV}
\end{figure}

The case of SB2 is more challenging because the spectroscopic variability is not merely a simple Doppler shift. We have simulated SB2 data using two PHOENIX spectra. We used the same solar-like spectrum as in the SB1 above as the spectrum of the primary component of the binary. For the secondary we used a spectrum corresponding to $T_\mathrm{eff}=5500\,\mathrm{K}$, $\log g = 4.5,$ and $[\element{Fe}/\element{H}]=0.0$. We shifted and blended the spectra, assuming a moderately eccentric ($e=0.3$) $\text{seven}$-day Keplerian orbit.  The orbital orientation was determined so that the maximum RV separation ($K_1+K_2$) would be $10\,\mathrm{km\,s}^{-1}$. In order to determine the individual semi-amplitudes $K_1$ and $K_2$, as well as the intensity ratio for combining the spectra, we used the masses and radii listed in PHOENIX, assuming the two stars are main-sequence stars. In total, we sampled the simulated orbit at $20$ epochs, with an S/N of $30$. Fig.~\ref{fig:SB2RV} presents the simulated primary and seconday RVs.

\begin{figure}
\includegraphics[width=\columnwidth,clip=true]{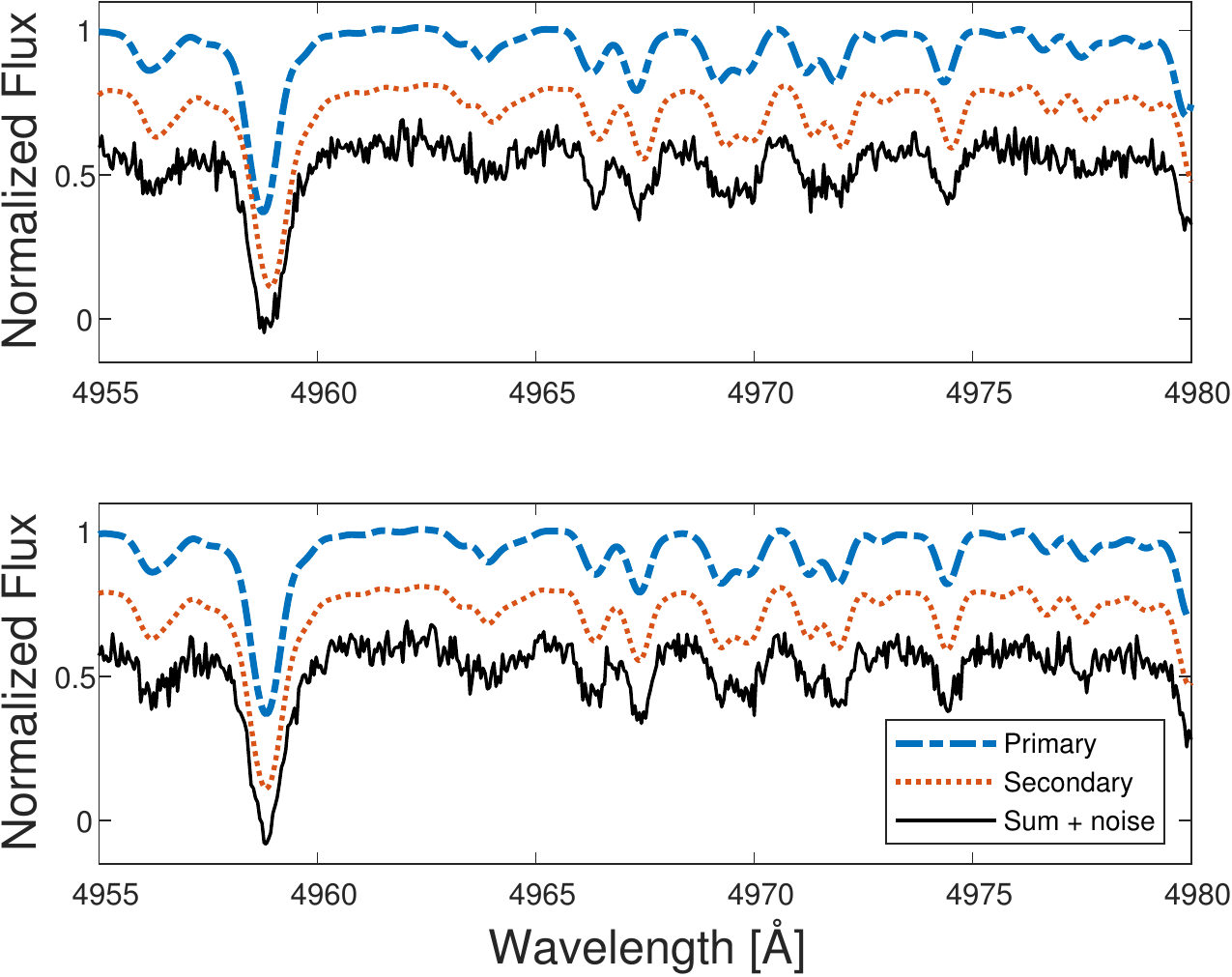}
\caption{Selected segment from two simulated SB2 spectra. The dashed blue lines represent the primary PHOENIX spectrum, and the dotted red lines represent the secondary. The solid black line is the combined and noised spectrum with an S/N of $30$. used for the simulation. The spectra are normalised to a continuum level of $1$. For clarity, a vertical offset of $0.2$ was introduced to separate the spectra. The upper panel shows the spectrum with the maximum RV separation, and the lower panel shows the spectrum with the smallest separation.}
\label{fig:SB2Spec}
\end{figure}

Fig.~\ref{fig:SB2Spec} demonstrates the challenge in this specific SB2 case. We show in the figure two of the $20$ spectra, at the largest and smallest RV separation. The figure focuses on the wavelength range $4955$\,--\,$4980$\,\AA, which includes the Fraunhofer iron c-line, at $4959.0$\,\AA\ (note that we did not convert PHOENIX spectra from vacuum to air wavelengths). The figure shows both components (with dashed blue and dotted red lines), and the composite noised spectrum (solid black line). For clarity we have introduced vertical offsets among the three spectra in each panel. The challenge is obvious: at a resolution $10000$ and S/N $30,$ it is practically impossible to distinguish the two components. The main effect of the varying RV separation seems to be a minute change in the width and depth of the composite spectral lines. 

\begin{figure}
\includegraphics[width=\columnwidth,clip=true]{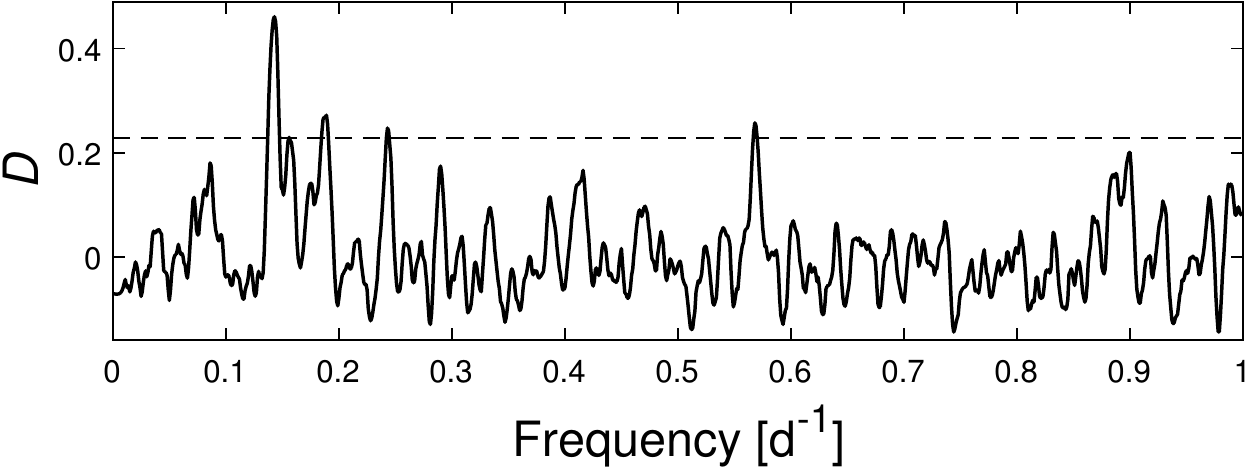}
\caption{USuRPER periodogram plot for the simulated SB2 case. The distance-correlation values of USuRPER are normalised and therefore unitless. The dashed line corresponds to an FAP level of $10^{-3}$, obtained by the permutation test procedure.}
\label{fig:SB2}
\end{figure}

Fig.~\ref{fig:SB2} presents the resulting USuRPER periodogram. In spite of the challenge posed by the low resolution, relatively low S/N and small RV separation, the maximum is obtained at a clear peak around the correct frequency, safely above the $10^{-3}$-FAP threshold. The new periodogram appears to perform reasonably well in this quite challenging case as well.

\subsection{Periodic temperature variability}
\label{subsec:ceph}

In addition to the examples above, we wished to test whether USuRPER is indeed also sensitive to other types of spectroscopic periodicities, not merely those related to periodic Doppler shifts. The periodic expansion and contraction phases of pulsating stars cause periodic Doppler shifts, but are also accompanied by cooling and heating. We therefore decided to simulate such periodic temperature changes, using the PHOENIX library, without the Doppler shift, so that the spectral features that change periodically would not be easily describable in a simple manner like Doppler shifts.

\begin{figure}
\includegraphics[width=\columnwidth,clip=true]{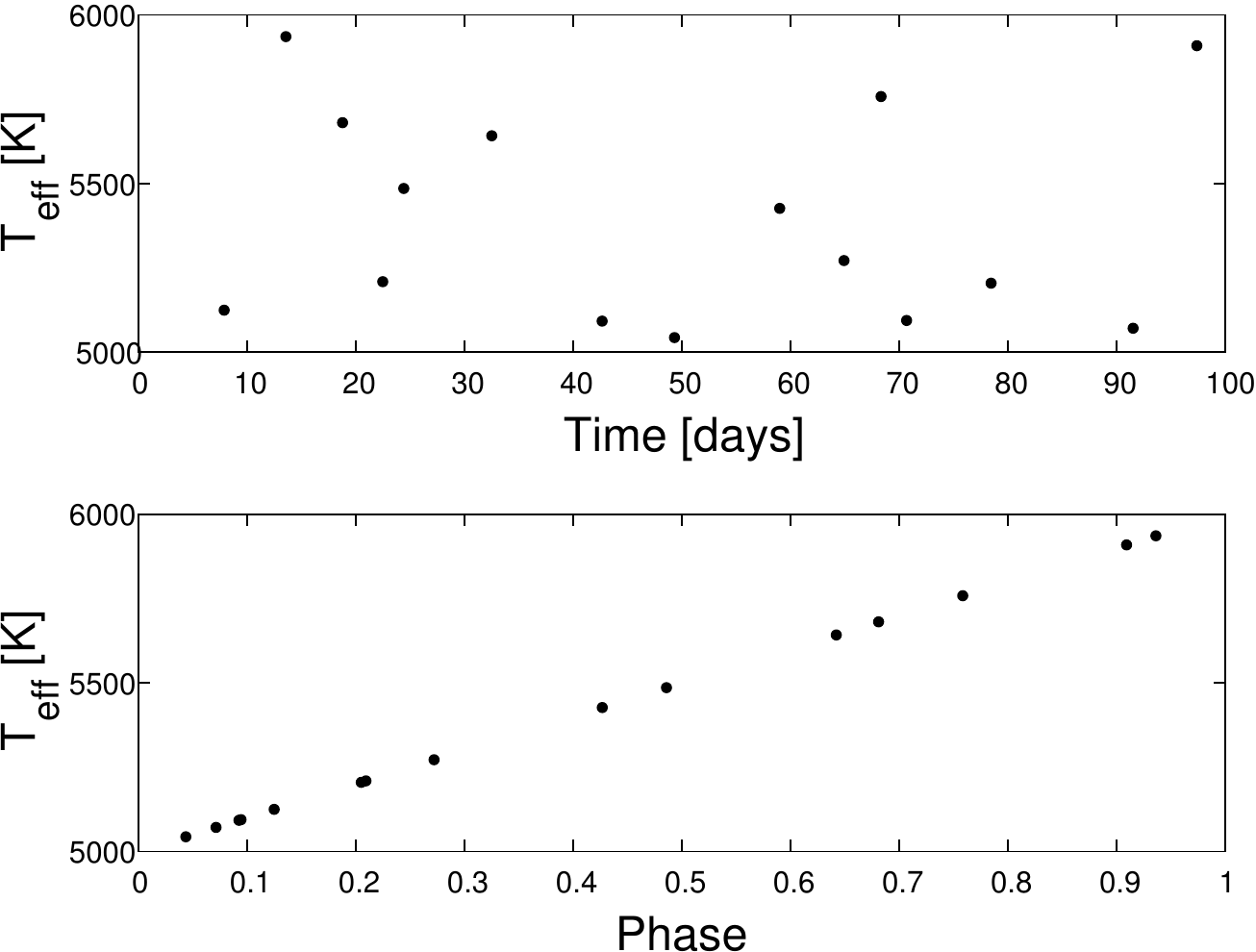}
\caption{Upper panel: Effective-temperature time series used in the periodic temperature variability simulation. Lower panel: Same effective-temperature time series phase-folded by the known $\text{seven}$-day period.}
\label{fig:CephTemp}
\end{figure}

\begin{figure}
\includegraphics[width=\columnwidth,clip=true]{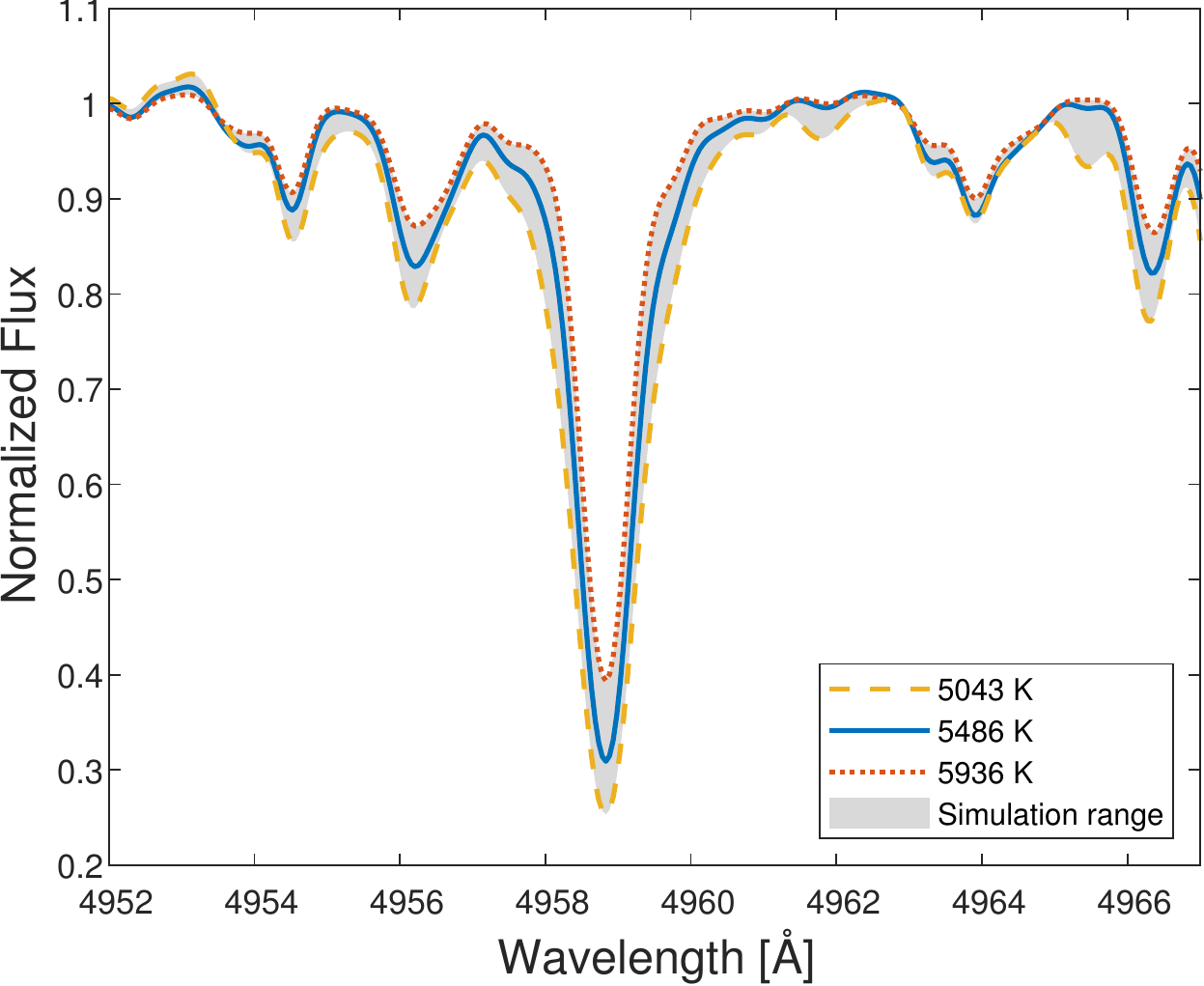}
\caption{Selected segment of the simulated spectra with periodic $T_\mathrm{eff}$ variability, before adding noise. The dashed yellow line corresponds to the spectrum with the lowest temperature ($5043$\,K) and the dotted red line to the highest temperature ($5936$\,K). The solid blue line represents a temperature in the middle ($5486$\,K). The shaded area represents the range between the spectra with the extreme temperatures.}
\label{fig:CephSpec}
\end{figure}

We simulated a saw-tooth effective-temperature variability, with $T_{\mathrm{eff}}$ varying between $5000\,\mathrm{K}$ and $6000\,\mathrm{K}$, and a period of $\text{seven}$~days by a simple linear interpolation over the PHOENIX temperature grid. This is a rough approximation to typical $T_{\mathrm{eff}}$ variability of classical Cepheids \citep[e.g.][]{Andetal2005}. We simulated $15$ random epochs, again over an interval of $100$ days, with an S/N of $30$ (Fig.~\ref{fig:CephTemp}). Fig.~\ref{fig:CephSpec} focuses on a narrow wavelength range of $4952$\,--\,$4967$\,\AA\ around the iron c-line and and shows how the spectrum changes as a result of the variable effective temperature (without the added noise). The dashed yellow line represents the spectrum of the lowest temperature simulated ($5043$\,K) and the dotted red line shows the highest temperature ($5936$\,K). A spectrum of a temperature in the middle ($5486$\,K) is also plotted with a blue line. The range of simulated temperatures is shaded in grey. The minute changes in the equivalent widths of the lines caused by the varying effective temperature are visible, without any bulk Doppler shift. Moreover, different lines behave quite differently, and might even exhibit different trends in equivalent width, as the temperature varies.

\begin{figure}
\includegraphics[width=\columnwidth,clip=true]{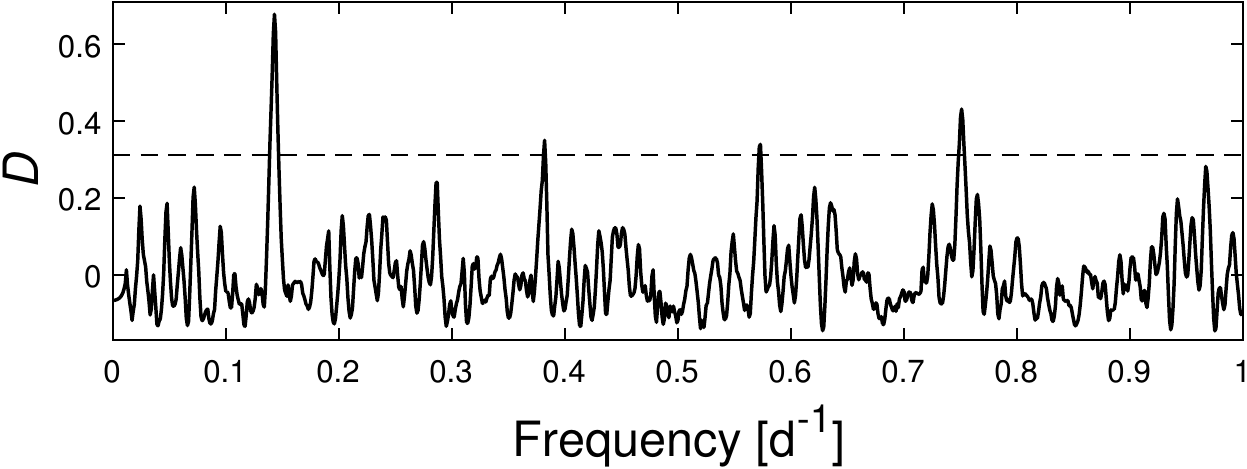}
\caption{USuRPER periodogram plot for the simulated temperature periodicity case. The distance-correlation values of USuRPER are normalised and therefore unitless. The dashed line corresponds to an FAP level of $10^{-3}$, obtained by the permutation test procedure.}
\label{fig:Ceph}
\end{figure}

Fig.~\ref{fig:Ceph} shows the result of the USuRPER periodogram applied to this dataset. In spite of the less favourable conditions, where there are fewer samples than the previous examples, and the S/N is not optimal, the peak at the correct period, much higher than the $10^{-3}$-FAP threshold, is evident, confirming that our novel periodogram performs well also in cases in which the periodicity is very different from simple Doppler shifts. 

It should be noted that in this specific simulation, a few lower spurious peaks appear to marginally cross the detection as well. Their frequencies seem to be around half-harmonics of the simulated frequency. This might be a random finding, but it should be further explored. In any case, the correct peak is definitely much more significant.

\subsection{Composite periodicity}
After we have demonstrated that USuRPER is sensitive to both RV and temperature periodic variability, it is interesting to test how it performs when presented with a composite type of periodicity, such as periodic RV variability combined with periodic temperature variability, with different periods. To this end, we again simulated a set of $50$ spectra. The simulated temperature variability resembled the one in Sect. \ref{subsec:ceph}, but with a period of $\text{five}$~days, whereas the RV variability was a sinusoidal variability similar to that in Sect. \ref{subsec:SB1}, with a period of $\text{three}$~days. White Gaussian noise was added at a level corresponding to an S/N of $100$.

\begin{figure}
\includegraphics[width=\columnwidth,clip=true]{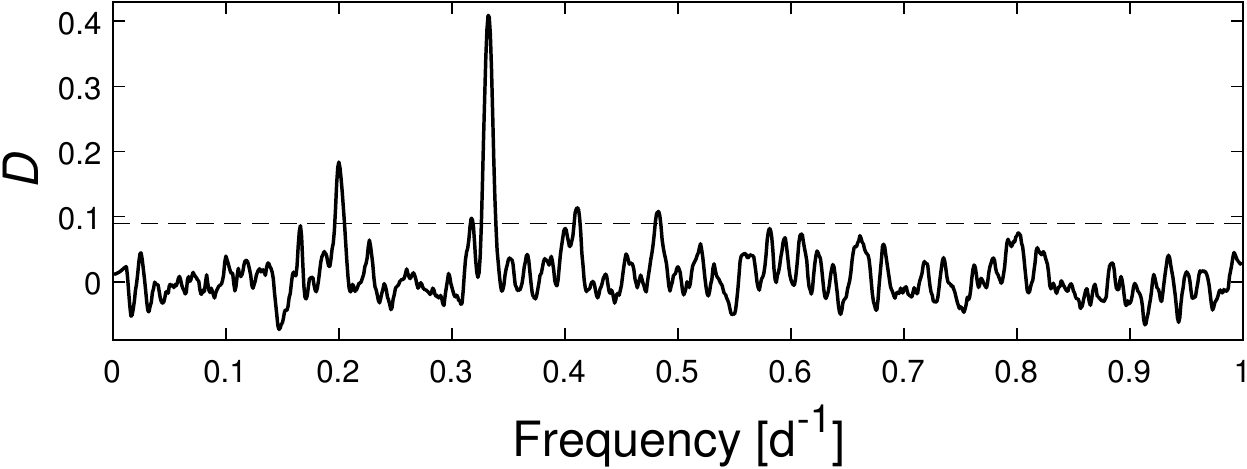}
\caption{USuRPER periodogram plot for the simulated composite temperature and RV periodicity case. The distance-correlation values of USuRPER are normalised and therefore unitless. The dashed line corresponds to an FAP level of $10^{-3}$, obtained by the permutation test procedure.}
\label{fig:CephSB1}
\end{figure}

The two corresponding peaks, at frequencies $1/3$ and $1/5\ \mathrm{d}^{-1}$, are clearly seen in the USuRPER periodogram of these data, in Fig.~\ref{fig:CephSB1}. They are both safely higher than the $10^{-3}$ significance threshold, but they are still not of the same prominence, however. This probably reflects the fact that the effects of temperature and RV periodicities, at the simulated amplitudes, do not have the same impact on the overall variability of the spectrum. Nevertheless, the presence of both peaks in the periodogram shows that they did not in some way interfere in a destructive fashion that would make them disappear. This serves to show that USuRPER can also be used for cases of multiple periodicities. The case of a temperature periodicity combined with an RV periodicity of a different period can be encountered\ in cases of Cepheids in spectroscopic binary stars \citep{Szaetal2013}, for instance.

\subsection{HD 115226}
\label{subsec:HD115226}

In the previous examples we have applied USuRPER on sequences of simulated spectra, which were obviously far better behaved than real-life data. We therefore looked for a publicly available real-life time-resolved spectroscopy dataset exhibiting spectral variability, preferably of a different type from those of the previous examples. We finally decided to test USuRPER on observed spectra of a known rapidly oscillating Ap (roAp) star.

Broadly speaking, roAp stars are stars that exhibit very short-period photometric or RV variations, with periods of the order of minutes \citep[e.g.][]{Kur1990}. \citet{Ryaetal2007} have further characterised the spectral variability of roAp stars by showing that absorption lines of some of the heavier chemical species (rare-element ions) perform periodic Doppler shifts, usually all with the same period, but not with the same amplitude or phase. This means that the overall spectrum shape changes periodically with a rather complicated pattern, which renders analysis by cross-correlation ineffective. Instead, the common approach is to analyse each individual line separately, measure its Doppler shift, and analyse its periodicity.

\citet{Kocetal2008} have observed the roAp star \object{HD\,115226} using HARPS \citep{Mayetal2003}. They obtained time-series spectroscopy of HD\,115226 including $102$ spectra during a time interval of $4.3$~hours, and performed a meticulous RV analysis of various absorption lines. The analysis yielded an estimated oscillation period of $10.87\pm 0.01$\,min. 

\begin{figure}
\includegraphics[width=\columnwidth,clip=true]{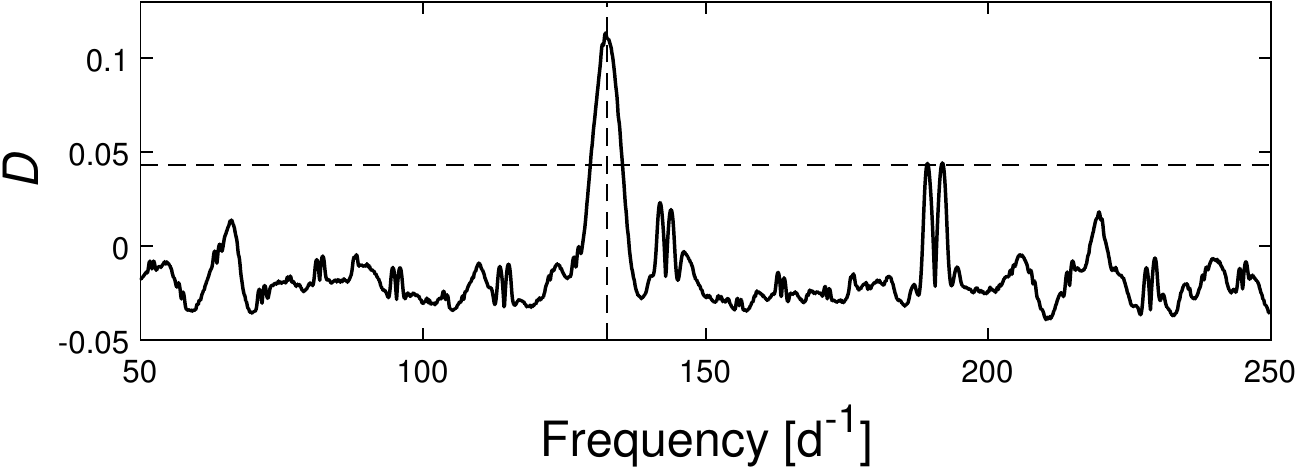}
\caption{USuRPER periodogram plot for the HARPS spectra of HD\,115226. The vertical dashed line represents the known period of $10.87$\,min \citep{Kocetal2008}. The distance-correlation values of USuRPER are normalised and therefore unitless. The horizontal dashed line corresponds to an FAP level of $10^{-3}$, obtained by the permutation test procedure.}
\label{fig:roAp}
\end{figure}

We downloaded the $102$ HARPS spectra, and applied USuRPER on this dataset. Based on table~3 of \citet{Kocetal2008}, we restricted the wavelength range we analysed to $4900$\,--\,$5150$\,\AA, where a few important \ion{Nd}{iii} lines are located. A wider wavelength range would have diluted the periodicity information because most of the spectral features in other wavelengths do not exhibit periodicity. Knowing that we searched for a phenomenon with a typical period of a few minutes, we ran USuRPER on a frequency range of $50$\,--\,$250\,\mathrm{d}^{-1}$, corresponding to a period range of $5.76$\,--\,$28.8\,\mathrm{min}$. Fig.~\ref{fig:roAp} shows the resulting periodogram. 

The obvious maximum is at a frequency of $132.3\,\mathrm{d}^{-1}$, corresponding to a period of $10.88\,\mathrm{min}$, in agreement with the period \citeauthor{Kocetal2008} had obtained by their individual-line analysis. A conventional error estimate for the period is difficult to estimate because it requires some modelling of the periodicity \citep[e.g.][]{Baletal1995}. However, some confidence interval can be estimated using the frequencies around the peak where the periodogram crosses the $10^{-3}$-FAP threshold. The resulting estimate of the period is $10.88\pm0.28\,\mathrm{min}$. The uncertainty is larger than the uncertainty reported by \citeauthor{Kocetal2008}, but this is expected because they used a specific known model for the periodicity, and also included more absorption lines in their analysis. In any case, the two period estimates perfectly agree within their error bars. The peak is also well above the $10^{-3}$-FAP significance threshold. Interestingly, two additional twin peaks are clearly seen around the frequency $190\,\mathrm{d}^{-1}$, but they barely reach the $10^{-3}$-FAP threshold, and are probably spurious.

\section{Conclusion}
\label{sec:conc}

The examples we have shown above attest to the wide potential of the USuRPER periodogram. We have shown that it performs well in cases of RV periodicities, composite SB2 spectra, and even complicated spectrum-shape patterns such as periodic temperature changes. We also demonstrated its performance in real-life cases of exotic variability such as roAp stars. We provide our Python implementation of USuRPER in the form of a public GitHub repository\footnote{USuRPER is available as part of the SPARTA package, at \url{https://github.com/SPARTA-dev/SPARTA}.}. 

In order to estimate the significance of peaks in the USuRPER periodogram, simple bootstrap-like permutation tests can be performed in which the time stamps of the individual spectra would be repeatedly randomly shuffled, in order to obtain the null distribution of the distance-correlation values under the assumption of no dependence.

Because USuRPER does not provide any further information about the nature of the periodicity, except for the period and its significance, it is essentially useful as an exploratory tool. Once a prominent peak appears in the periodogram, further analysis is required in order to tell whether the observed object is a binary star (or exoplanet), a pulsating star, or maybe some other type of periodicity we did not encounter before. 

An important application of USuRPER can be, for example, to use it in the analysis of the RVS, BP and RP spectra of \textit{Gaia} \citep{GaiCol2016}, or other large spectroscopic surveys with potentially multiple visits per object, for instance, APOGEE \citep{Majetal2017} or LAMOST \citep{Cuietal2012}. Another interesting application might be the study of periodic stellar variability patterns that might interfere with the detection of exoplanets through minute Keplerian RV variations \citep{Boietal2011}. 

The USuRPER periodogram offers a completely new approach to study astronomical spectra. An approach that may very well pave the way to new discoveries and insights, potentially ones that cannot be discovered in any other way.

\begin{acknowledgements}
We thank the anonymous referee for their wise comments that helped to improve the manuscript.
We are grateful to Aviad Panahi for patiently examining our USuRPER code implementation, and to Dolev Bashi for reviewing and commenting on an early version of the manuscript. This research was supported by the ISRAEL SCIENCE FOUNDATION (grant No. 848/16). We also acknowledge partial support by the Ministry of Science, Technology and Space, Israel. The research is partly based on observations collected at the European Southern Observatory, La Silla, Chile (ESO program 079.D-0118). The analyses done for this paper made use of the code packages: Astropy \citep{Ast2013,Ast2018}, NumPy \citep{vanetal2011}, SciPy \citep{Viretal2020}, and PyAstronomy \citep{Czeetal2019}.
\end{acknowledgements}

\end{document}